# Governing Copyright Responsibly in the Era of Generative AI

Dmitry Kochetkov[1,2]
[1]Centre for Science and Technology Studies (CWTS), Leiden University, Leiden, the Netherlands
[2]RUDN University, Moscow, Russia

E-mail: d.kochetkov@cwts.leidenuniv.nl; kochetkov-dm@rudn.ru
ORCID: 0000-0001-7890-7532

**Abstract.** Large language models are trained on collections that may include scholarly works, while authors generally lack work-level information about whether and how their works were used. This paper asks how copyright governance frames and operationalises the protection of such authors, and what a responsible alternative would require. Forty legislative, judicial and soft-law instruments issued between 2019 and 2026 were coded deductively against a framework derived from Responsible Research and Innovation theory, yielding 1,337 category assignments across 835 segments. The European Union and the United States are the primary cases; the United Kingdom, China and international bodies serve as comparators. Normative orientation and the distribution of responsibility dominate the corpus, while reflexivity is the least represented category and reaches its minimum in the US sub-corpus. Both primary jurisdictions frame protection procedurally rather than as an entitlement, and no instrument in the corpus creates a general author-facing attribution duty[1] for the use of works in AI training. From the coded gaps, the paper derives a six-instrument governance design and a capacity-based allocation of responsibility among developers, publishers, collecting societies, funders, institutions, regulators, and individual authors.

## 1. Introduction

Large language models are trained on collections that may include scholarly works, yet authors generally have no work-level means of knowing whether, or on what basis, their works were used. The evidence that would establish such use is held by developers and data intermediaries, not by the authors whose rights are at stake.

The intersection of AI training and copyright law has generated substantial legal scholarship examining how different jurisdictions approach these challenges. Sag & Yu (2024) identify an emerging international equilibrium where countries recognise text and data mining as potentially valuable while maintaining some copyright protections, driven by the centrality of the idea-expression distinction, global AI competition, and reform convergence. Chopra et al. (2025) offer comparative analysis of how courts and policymakers in multiple jurisdictions address generative AI's impact on copyright and personality rights, highlighting divergent regulatory approaches between US litigation-based doctrines, EU statutory frameworks, and emerging jurisprudence in other legal systems. The European Parliament 2025 study on generative AI and copyright examines structural risks to European creative economy, calling for harmonised opt-out mechanisms and equitable licensing models (Lucchi, 2025). Senftleben (2025) proposes policy pathways to balance innovation with creator protection. Quintais (2025) provides a detailed critique of the EU AI Act's legal ambiguities in terms of copyright.

[1] A "general author-facing attribution duty" denotes an obligation, owed to authors, to acknowledge the use of their works in AI training; it is distinct from moral rights of attribution, which concern the treatment of the work itself rather than training-data practices, and from output-side citation rules, which govern the presentation of generated results rather than the ingestion of source works.

This study draws on the Responsible Research and Innovation framework (RRI), which crystallised as a distinct policy concept in European research governance following a European Commission workshop in May 2011 (Owen et al., 2012). Von Schomberg (2011) defined RRI as *"a transparent, interactive process by which societal actors and innovators become mutually responsive to each other with a view on the (ethical) acceptability, sustainability and societal desirability of the innovation process and its marketable products"* (p. 9). Stilgoe et al. (2013) operationalised this definition through four process dimensions: Anticipation, Inclusion, Reflexivity, and Responsiveness (AIRR). These dimensions have since been widely adopted as an analytical framework for evaluating governance arrangements in research-intensive fields.

Owen et al. (2021) revisited the RRI framework a decade after its emergence. They acknowledged that integration of RRI into Horizon 2020 produced substantial institutional change in some domains while leaving others, notably industrial and commercial innovations, largely unaddressed. Buhmann & Fieseler (2023) argue that responsible innovation in AI requires distributed deliberation that includes expert, corporate, and civil-society actors. This paper explores a specific governance failure: existing copyright rules and open licences do not consistently give scholarly authors an effective means to identify, authorise, or obtain remuneration for the use of their works in AI training.

A central theoretical claim is that the governance failures documented in this study support what Von Schomberg (2011) presented as the Collingridge dilemma of ICT governance: "*ethical issues could be easily addressed early on during technology design and development whereas in this initial stage the development of the technology is difficult to predict. Once the social and ethical consequences become clearer, the development of technology is often far advanced and its trajectory is difficult to change."* Von Schomberg argues that the appropriate response to the Collingridge dilemma is early societal intervention, structured technology assessment, and stakeholder engagement during the design and deployment stage. These are precisely the mechanisms that current copyright governance failed to provide.

This paper examines how the existing copyright frameworks fail to govern the use of scholarly works in LLM training and proposes directions for reforms aimed at protecting authors' rights while sustaining the open-science ecosystem.

*RQ1:* How do US and EU governance instruments frame and operationalise protection of authors whose works are used for LLM training?

*RQ2a:* What governance instruments can address the identified gaps in accordance with the RRI framework?

*RQ2b:* What roles and responsibilities should key stakeholders assume in design and enactment of these instruments in line with the distribution of responsibility?

The paper contributes to responsible-innovation scholarship in three ways. First, the paper provides a systematic empirical application of the AIRR framework to a cross-jurisdictional legal corpus. The codebook in Section 3 converts the four process dimensions, together with four copyright-specific categories, into observable properties of governance instruments, and the coding shows where each dimension is carried in the legal architecture and where it is absent. Second, it identifies reflexivity as least represented in the US sub-corpus, a jurisdiction hosting many frontier-model developers. Third, it shows that the AIRR categories can function as design requirements and not only as evaluative criteria: Section 6 uses the coded categories to derive a six-instrument governance design and a capacity-based allocation of stakeholder responsibility from the documented gaps. For RRI research, the study turns the AIRR framework from a set of process aspirations into observable properties of governance instruments; for copyright policy, it identifies the governance functions that no current instrument performs.

An earlier preprint (Kochetkov, 2025) offered an interpretative review of AI-related copyright regulation and case law. The present article is a separate study based on the deductive RRI-driven qualitative content analysis.

## 2. State of the Art

### *2.1. The Emergence of the RRI Framework*

Owen et al. (2021) traced the evolution of the RRI concept since its emergence in 2011, through institutionalisation as "five keys"[2] to its partial absorption into Open Science under Horizon 2020. Woolley et al. (2024) summarise the critiques of this trajectory: RRI emerged in north-western Europe and travels badly beyond its region of origin, remained confined to public discourse and the academic community rather than commercially driven innovation, and has been partially replaced by Open Science, one of its own keys.

Buhmann & Fieseler (2023) argued that AI governance poses challenges qualitatively different from those in other complex fields such as bioethics. The key difference is epistemic opacity, under which internal mechanisms, evidence base and long-term consequences of AI deployment are not reliably known even by the system's own developers. Their taxonomy distinguishes strategic opacity, which is the developers' intention to keep the underlying mechanisms "black-boxed" so as to avoid accountability, from expert opacity, the gap between what AI specialists and what citizens and policymakers can comprehend, and from true opacity, which "arises from the ways in which AIs are developed and evolve as emergent phenomena in practice".

Dual Use Research of Concern (DURC) refers to studies whose knowledge, products or technologies "could be directly diverted" to pose a threat to public health, agriculture, wildlife, the environment, or national security (*Biotechnology Research in an Age of Terrorism*, 2004). Grinbaum & Adomaitis (2024) adapted the concept to LLM development and deployment, arguing that beyond mere misuse LLMs demonstrate a potential for malevolent or negligent use and could produce novel types of harm.

In terms of copyright, the novel harm is knowledge asymmetry: Open Science, designed to reduce inequality between researchers from rich and poor countries, was repurposed by LLM developers to produce new inequality between those who own the model and those who do not. In this study, dual use is coded only where a governance mechanism, dataset practice, or technical capability can serve a legitimate function while also facilitating circumvention, untraceable ingestion, or other copyright-related harm.

### *2.2. The Difference Between the US and EU Copyright Legal Frameworks*

Carlini et al. (2026) argue that the EU and the United States are following sharply divergent paths on copyright and AI training, contrasting the EU's statutory text-and-data-mining exception framework with the US reliance on case-by-case fair use adjudication.

The central concept in the EU regime is text and data mining (TDM). Article 2(2) of the Directive on Copyright and Related Rights in the Digital Single Market (2019) (DSM Directive) defines it as "any automated analytical technique aimed at analysing text and data in digital form in order to generate information which includes but is not limited to patterns, trends and correlations." Articles 3 and 4 of the same Directive establish two TDM exceptions, distinguished by the actors who may rely on them and by the purpose of the mining. Article 3 covers reproductions and extractions made by research organisations and cultural heritage institutions for the purposes of scientific research on works to which they have lawful access. Article 4 establishes a general exception for reproductions and extractions of lawfully accessed works, subject to rights reservation by rightsholders, including through "machine-readable means in the case of content made publicly available online."

[2] The European Commission initially defined six keys (gender equality, ethics, public engagement, science education, open access, and governance); however, the "governance" key was dropped in practice.

The EU AI Act (Regulation (EU) 2024/1689 of the European Parliament and of the Council of 13 June 2024, 2024) builds on the DSM Directive by seeking to operationalise the opt-out mechanism of Article 4(3). Article 53(1) imposes two copyright obligations on providers of general-purpose AI (GPAI) models: Article 53(1)(c) to put in place a policy to comply with EU copyright law, including a mechanism for identifying and respecting reservations of rights, and Article 53(1)(d) to draw up and make publicly available a sufficiently detailed summary of the content used for training. Article 53(1)(d) carries the transparency function of the regime. Rightsholders who wish to establish whether their reservations have been respected have no independent means of inspecting a training corpus, so the published summary is the only routine source of information available to them. The obligation is procedural in form: it requires documentation calibrated to regulatory oversight rather than disclosure matched to the informational needs of those who would act on it.

US law contains no equivalent statutory exception for AI training. The question falls instead to the fair use doctrine, which requires assessment under four statutory factors: the purpose and character of the use, the nature of the copyrighted work, the amount and substantiality of the portion used, and the effect of the use on the potential market for the work. In May 2025, the US Copyright Office released Part 3 of its Report on Copyright and Artificial Intelligence, which addresses the application of these factors to generative AI training (*Copyright and Artificial Intelligence, Part 3: Generative AI Training*, 2025). Its principal conclusion is that such training does not automatically qualify as fair use and that each case requires analysis under all four factors.

Courts applying that test to AI training have reached divided outcomes. For instance, in *Bartz v. Anthropic PBC, No. 3:24-cv-05417* (N.D. Cal. 2025), the court held on partial summary judgment that training on lawfully obtained books was fair use, while the acquisition and retention of pirated copies in its central library was not. In *Thomson Reuters Enterprise Centre GmbH v. Ross Intelligence Inc., No. 1:20-cv-613-SB* (D. Del. 2025), the court rejected the fair use defence on summary judgment, finding the use commercial and not transformative and recognising a licensing market for training data. The two rulings differ in subject matter and procedural posture, and the doctrine remains unsettled: the same four factors have so far yielded different results depending on data provenance, market proximity, and the type of system at issue.

The EU's statutory architecture produces uniform ex ante obligations documented in advance by all providers, whereas the US adjudicative architecture produces precise but fragmentary rules available only after litigation, and this difference in timing and generality yields distinct governance profiles.

### *2.3. Attribution and Credit in Scholarly Communication*

Attribution identifies the origin of a scholarly contribution and traces credit to its author. Credit is "the currency of the realm" of science (Dasgupta & David, 1994, as cited in Flier, 2019): without it, scientists cannot secure appointments, promotions, or funding, and priority controversies "have long been frequent, harsh and ugly" (Merton, 1969, as cited in Flier, 2019). Since the mid-19th century, publication in a scholarly journal has been the core mechanism for claiming credit and priority (Flier, 2019), and the citation, "an explicit, intentional expression of influence" (Bollen, 2012), is its basic unit. Attribution is therefore both a normative instrument of the reward system and the epistemic infrastructure that lets readers trace claims to sources.

That infrastructure was under strain before generative AI arrived. Larger interdisciplinary teams and proliferating journals have made credit allocation harder and authorship disputes common (Flier, 2019), while assessment has remained "stuck in the paper era" even as scholarly communication moved online (Bollen, 2012).

Generative AI weakens the citation link on which the reward system depends. More than 30% of responses by leading search-enabled LLMs provided no attribution whatsoever (Hardinges & Bejan, 2026), and these gaps "result from design choices, not simply technical limitations" (Strauss et al., 2025). Where attribution is provided, it does not reliably carry value back to the source. Attribution creates "pathways for traffic and financial value to flow" to creators, yet "a creator can be attributed and still lose the traffic that funded their work" (Hardinges & Bejan, 2026).

In the terms of this study, attribution is therefore a governance issue rather than a technical function of publishing, and it bears on three of the coding categories. It is a matter of transparency and traceability, because credit cannot be claimed for a use that cannot be observed. It is a matter of the distribution of responsibility, because the duty to attribute is divided among authors, publishers, and model developers without settled allocation. It is also a matter of inclusion, because the capacity to enforce attribution is unevenly distributed between large rightsholders and individual scholars. Sections 4.2 and 4.3 trace how far existing instruments recognise these dimensions.

## 3. Data and Methods

### *3.1. Analytical Framework*

This study examines copyright governance for AI training in scholarly communication employing Responsible Research and Innovation (RRI) as its analytical framework. RRI took shape in European research policy as a corrective to retrospective, risk-based regulation of emerging technologies, foregrounding upstream societal engagement and shared responsibility (Owen et al., 2012; Stilgoe et al., 2013; Von Schomberg, 2013).

For qualitative content analysis, I use the AIRR framework proposed by Stilgoe et al. (2013), who defined responsible innovation as collective stewardship of science and technology in the present. AIRR supplies four top-level deductive categories for coding governance texts. It captures forward-looking governance capacities (anticipating impacts, adapting to emerging concerns) alongside the procedural conditions of legitimacy (inclusive problem framing and reflexive examination of assumptions). Because the RRI literature treats responsibility as distributed across innovation systems rather than reducible to ex post liability (Grinbaum & Adomaitis, 2024; Von Schomberg, 2013), the codebook also records how documents allocate duties and justify them in normative terms.

The coding framework therefore comprises the four AIRR categories together with three cross-cutting categories relevant to copyright governance in AI training: normative orientation, distribution of responsibility, and transparency and traceability. A further cross-cutting category, epistemic opacity and dual use, is applied where opacity constrains transparency, accountability, or stakeholder engagement (Buhmann & Fieseler, 2023; Grinbaum & Adomaitis, 2024). Whereas AIRR provides a core deductive structure, additional transverse categories capture AI/copyright specific governance properties not fully represented in AIRR.

The codebook is available in online supplement on Zenodo at https://doi.org/10.5281/zenodo.20555375 (Kochetkov, 2026).

*3.2. Data Sources*

The corpus draws on three types of primary sources: legislative and policy documents, judicial decisions, and international soft-law instruments. Selection was purposive: a document was included if it directly addresses the governance of AI training, text and data mining (TDM), or copyright protection for training data. The analysis focuses on the US and EU, while the UK and China are included as secondary comparators (Fig. 1). The temporal scope runs from 2019 to June 2026, bounded at the early end by the entry into force of the DSM Directive (Directive (EU) 2019/790 on Copyright and Related Rights in the Digital Single Market, 2019) and at the late end by the *CMA Publisher Conduct Requirement: Google's General Search Services (Final Decision)* (2026), which post-dates the draft agenda of the WIPO SCCR/48 Session (Geneva, May 18–22, 2026) (Standing Committee on Copyright and Related Rights, 2026).

Full citations for all corpus documents, including official references, dates, legal status, and source links, are provided in the online supplement (Kochetkov, 2026).

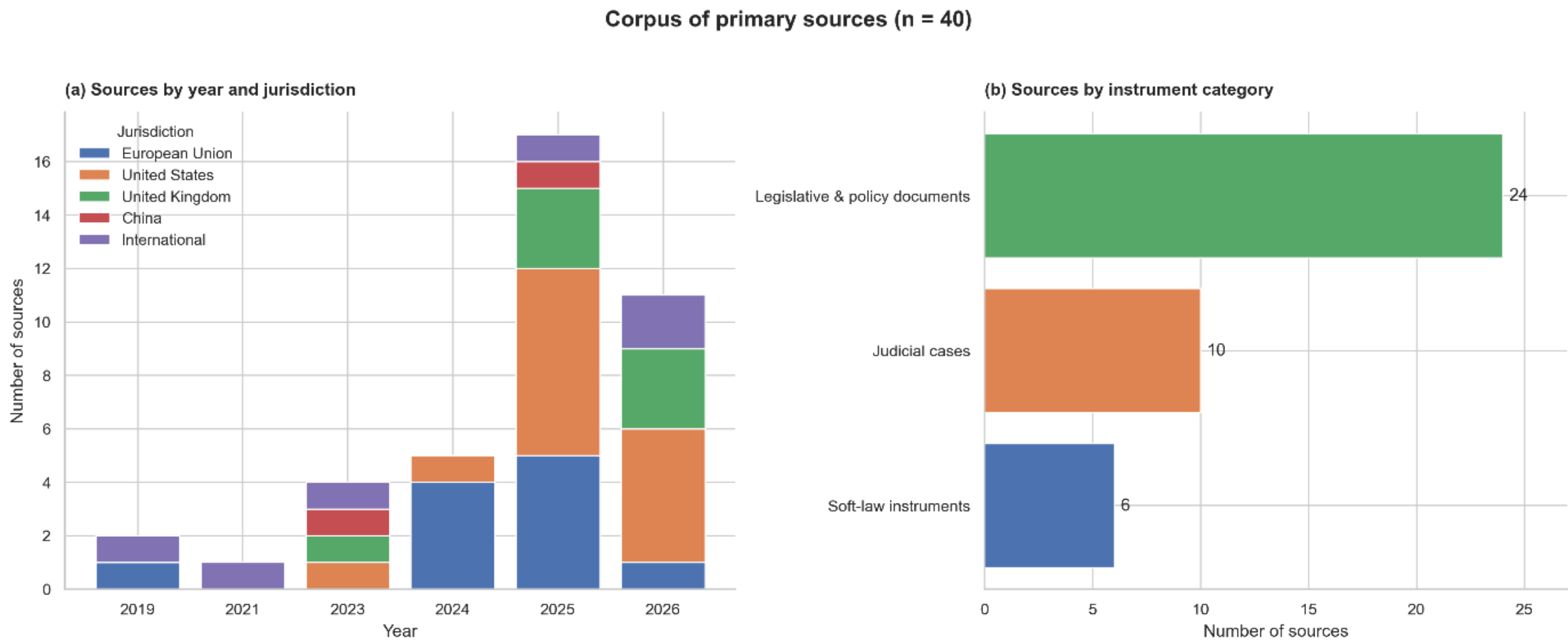

**Fig. 1.** Scope, jurisdiction, and temporal boundaries of the document corpus.

Two academic/policy documents are used as secondary sources to contextualise the primary corpus: the European Parliament study on Generative AI and Copyright (Lucchi, 2025) and Spichtinger's (2026) analysis of Creative Commons licensing in the context of AI training. These texts are not treated as governance documents subject to AIRR coding; they are cited in the analytical narrative to supply legal and contextual background. The same applies to the CLA/ALCS/PLS Collective Licensing Scheme for AI Training (UK, under development as of 2026), which would assign remuneration duties to AI developers through a collective organisation, partly addressing the limitations of individual opt-out (*PLS and ALCS Agree to Development of Pioneering CLA Generative AI Licence*, 2026).

The study is delimited in two respects. The corpus comprises copyright instruments and adjacent AI-governance instruments containing provisions or claims relevant to copyright, training data, attribution, or responsibility. Data-governance legislation, including the GDPR, the Data Governance Act, the Data Act, and the Digital Services Act, is excluded because these instruments regulate personal and non-personal data rather than protected subject matter. The analysis is further confined to the input side of generative AI, that is, the use of copyrighted works as training data. Output-side questions, including the protectability of AI-generated content and liability for infringing outputs, are not addressed. Where inference-time retrieval, retrieval-augmented generation, output attribution, or litigation discovery is discussed below, it is treated as boundary evidence bearing on input-side governance rather than as an object of analysis in its own right.

### *3.3. Data Analysis*

Doctrinal legal analysis typically proceeds by identifying leading cases and flagship statutes and reading them for their authoritative meaning. This study adopts a different approach. The corpus was analysed through systematic qualitative content analysis in Atlas.ti (version 26), without automated coding or other AI-driven functions. Systematic qualitative content analysis treats a corpus of governance documents as empirical data rather than as authoritative text, following Brook (2022) and Salehijam (2018). The aim is not to determine what the law means in a legally binding sense, but to describe what the law does: which governance problems it foregrounds, which actors it assigns responsibility to, and which questions it leaves unaddressed. Coding makes these patterns explicit and comparable across documents and jurisdictions.

The analysis is deductive rather than inductive. Coding proceeded in two stages. In the first stage, I applied the eight deductive categories from the codebook. In the second stage, I reviewed all coded segments within each category to identify patterns across documents and jurisdictions, noting convergences, contradictions, and gaps. The codebook was derived from the RRI literature before any engagement with the corpus, and the categories were held constant throughout coding. This follows the logic of deductive, theory-guided qualitative content analysis described by Mayring (2014).

Each document was imported and coded at the level of the meaning unit (sentence, paragraph, or provision) expressing a distinct claim about AI training, copyright, licensing, attribution, or governance. Meaning units were preferred to mechanical sentence-level coding because legal language is often compressed: a single provision may state an obligation, identify its addressee, and justify it normatively within one syntactic unit. Meaning units may receive multiple codes, since AIRR dimensions often coincide; a machine-readable opt-out standard, for instance, simultaneously concerns inclusion, responsiveness, and transparency.

The study was conducted by a single researcher. Although statutory wording constrains the range of plausible readings, assigning meaning units to theoretically derived categories remains interpretive. Analytical rigour was supported by an ex ante codebook held constant throughout coding, citation of coded passages in the Results, Atlas.ti quotation and code-frequency checks, and publication of the coded dataset. No inter-coder reliability test or delayed recoding was performed; accordingly, the category boundaries represent one informed and auditable reading of the corpus.

## 4. Results

### *4.1 Coding distributions and corpus coverage*

Coding across 40 primary sources produced 835 coded segments with 1,337 category assignments (mean: 1.60 codes per segment) (Kochetkov, 2026), indicating that governance statements in this corpus regularly implicate more than one RRI dimension. Across the full corpus, reflexivity accounted for 79 of 1,337 category assignments (5.9%), the smallest share among the eight categories.

Table 1 sets out governance profiles against the eight categories of the coding framework. The EU profile (11 documents, 455 assignments) is the more evenly loaded of the two primary jurisdictions. Normative orientation (24.4%) and distribution of responsibility (20.0%) lead, as they do corpus-wide, but transparency and traceability (13.2%) and inclusion (12.1%) also carry substantial weight, and the four AIRR process dimensions together account for 33.6% of EU assignments. Two categories are less represented. Anticipation registers 5.5%, the lowest jurisdictional share, a pattern consistent with the EU instruments' emphasis on specifying obligations rather than forecasting impacts. Reflexivity registers 5.7% and is concentrated in few instruments.

The US profile (14 documents, 333 assignments) is the most skewed in the corpus. Normative orientation (32.4%) and distribution of responsibility (22.5%) together take 54.9% of assignments, while the AIRR process dimensions take only 24.3%. Reflexivity falls to 2.1% (n = 7), the lowest non-zero value for any category in any jurisdiction. The pattern is consistent with an adjudicative architecture in which the governance work is performed elsewhere.

The UK is treated here as a comparator rather than a primary case. The comparison is closer than this designation suggests: 357 assignments from seven documents approach the EU's 455 from eleven, so the UK functions as a near-equal comparator rather than a small-scale control. Its sub-corpus (7 documents, 357 assignments) produces the most balanced profile in the corpus, with the narrowest observed range of category shares (9.2% to 23.2%), and the highest shares for anticipation (11.8%) and reflexivity (9.2%). The composition of the UK sub-corpus explains this: it is dominated by the Report and Impact Assessment published under sections 136 and 137 of the Data Use and Access Act 2025, documents whose genre is deliberative rather than prescriptive.

China is represented by two instruments (71 assignments), so its profile is indicative rather than comparable in statistical terms. Copyright appears as a residual constraint on otherwise lawful conduct rather than as an object of governance design. Providers must "respect intellectual property rights and commercial ethics" and must "use data and foundational models that have lawful sources." The Measures add that "where intellectual property rights are involved, the intellectual property rights that are lawfully enjoyed by others must not be infringed." Neither instrument establishes a mining exception, a rights reservation mechanism, or a remuneration route, so the questions that shape governance discussion in the EU and US do not arise.

The international group (6 documents, 121 assignments) records the highest combined share of the four AIRR dimensions (48.8%), approaching half of all assignments. It records the corpus's highest shares for transparency and traceability (17.4%), inclusion (14.9%), and responsiveness (14.0%), and the corpus's lowest share for distribution of responsibility (18.2%). It records no epistemic opacity coding at all (0.0%). The group that names the coordination problem most clearly is the group least able to act on it: its duties are addressed to states, and none of its instruments is binding.

Table 1. Distribution of coded categories by jurisdiction

| **Category** | **European Union (11 documents; n = 455)** | **United States (14 documents; n = 333)** | **United Kingdom (7 documents; n = 357)** | **China (2 documents; n = 71)** | **International instruments (6 documents; n = 121)** |
|---|---|---|---|---|---|
| Anticipation | 25 (5.5%) | 24 (7.2%) | 42 (11.8%) | 6 (8.5%) | 14 (11.6%) |
| Inclusion | 55 (12.1%) | 29 (8.7%) | 36 (10.1%) | 9 (12.7%) | 18 (14.9%) |
| Reflexivity | 26 (5.7%) | 7 (2.1%) | 33 (9.2%) | 3 (4.2%) | 10 (8.3%) |
| Responsiveness | 47 (10.3%) | 21 (6.3%) | 33 (9.2%) | 9 (12.7%) | 17 (14.0%) |
| Normative orientation | 111 (24.4%) | 108 (32.4%) | 48 (13.4%) | 20 (28.2%) | 19 (15.7%) |
| Distribution of responsibility | 91 (20.0%) | 75 (22.5%) | 83 (23.2%) | 13 (18.3%) | 22 (18.2%) |
| Transparency and traceability | 60 (13.2%) | 46 (13.8%) | 44 (12.3%) | 7 (9.9%) | 21 (17.4%) |
| Epistemic opacity and dual use | 40 (8.8%) | 23 (6.9%) | 38 (10.6%) | 4 (5.6%) | 0 (0.0%) |

Note. Coding unit: the meaning unit (sentence, paragraph, or provision). Denominators: category assignments within each jurisdiction (column totals). Percentages are within-jurisdiction shares of assignments and may differ

from 100 only through rounding; segments may receive multiple codes. Jurisdiction groups are mutually exclusive. Shares describe coded assignments in a purposive corpus, not the prevalence or quality of governance in the underlying legal systems; the China group comprises two documents and the international group six heterogeneous instruments, so both profiles are indicative.

Normalised within each jurisdiction, the profiles diverge (Fig. 2): the United States records the highest share of normative orientation (32.4%) and the second-highest share of distribution of responsibility (22.5%), while international instruments record the highest shares of transparency and traceability (17.4%), inclusion (14.9%), and responsiveness (14.0%), together with no epistemic-opacity and dual-use coding (0.0%). The United Kingdom records the highest reflexivity share (9.2%) and the most balanced profile across the eight categories.

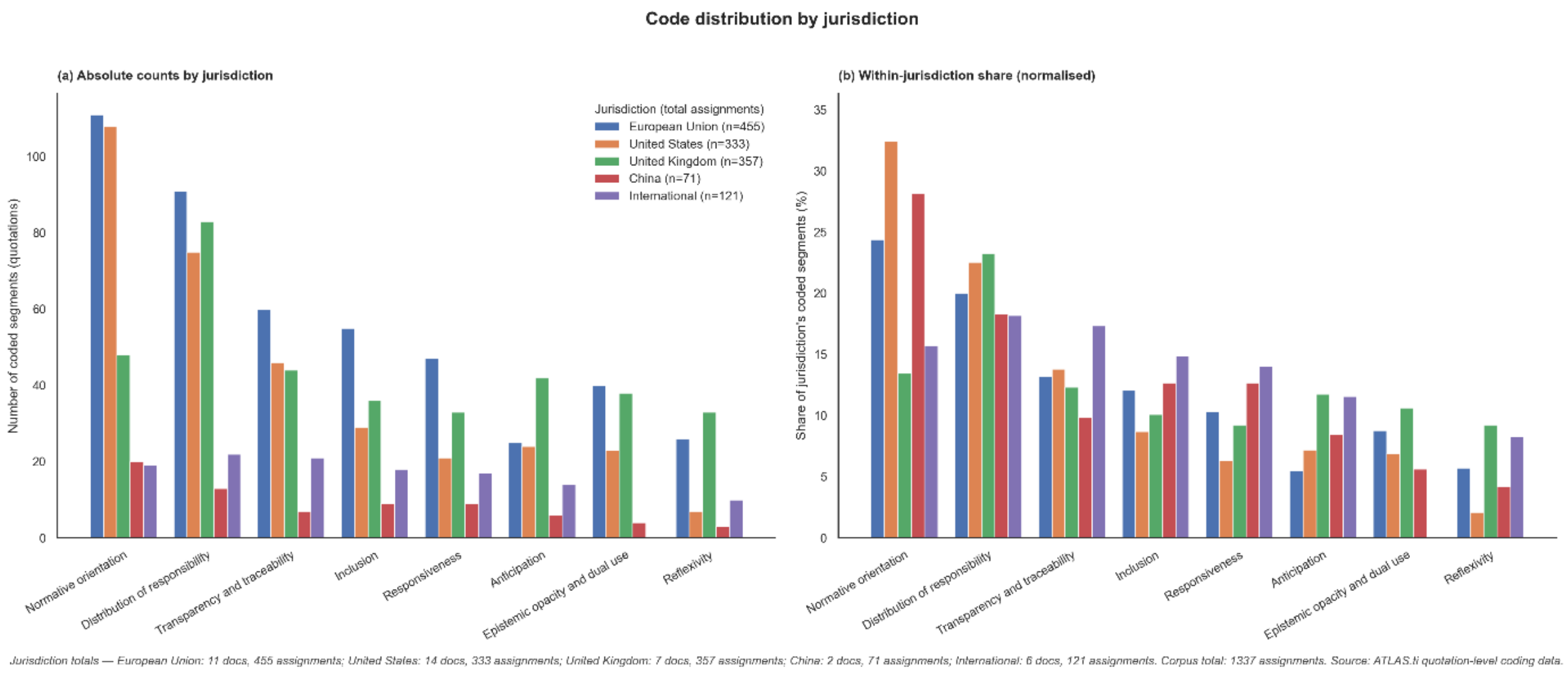

**Fig. 2.** Code distribution by jurisdiction.

The most frequent off-diagonal pair is normative orientation and distribution of responsibility (n = 100; Fig. 3): duty-allocation statements are routinely accompanied by normative justification. Transparency and traceability co-occurs often with distribution of responsibility (n = 68) but rarely with reflexivity (n = 10) or anticipation (n = 12). This pattern suggests that disclosure is more often framed as an allocated compliance duty than as a mechanism for anticipatory or reflexive learning.

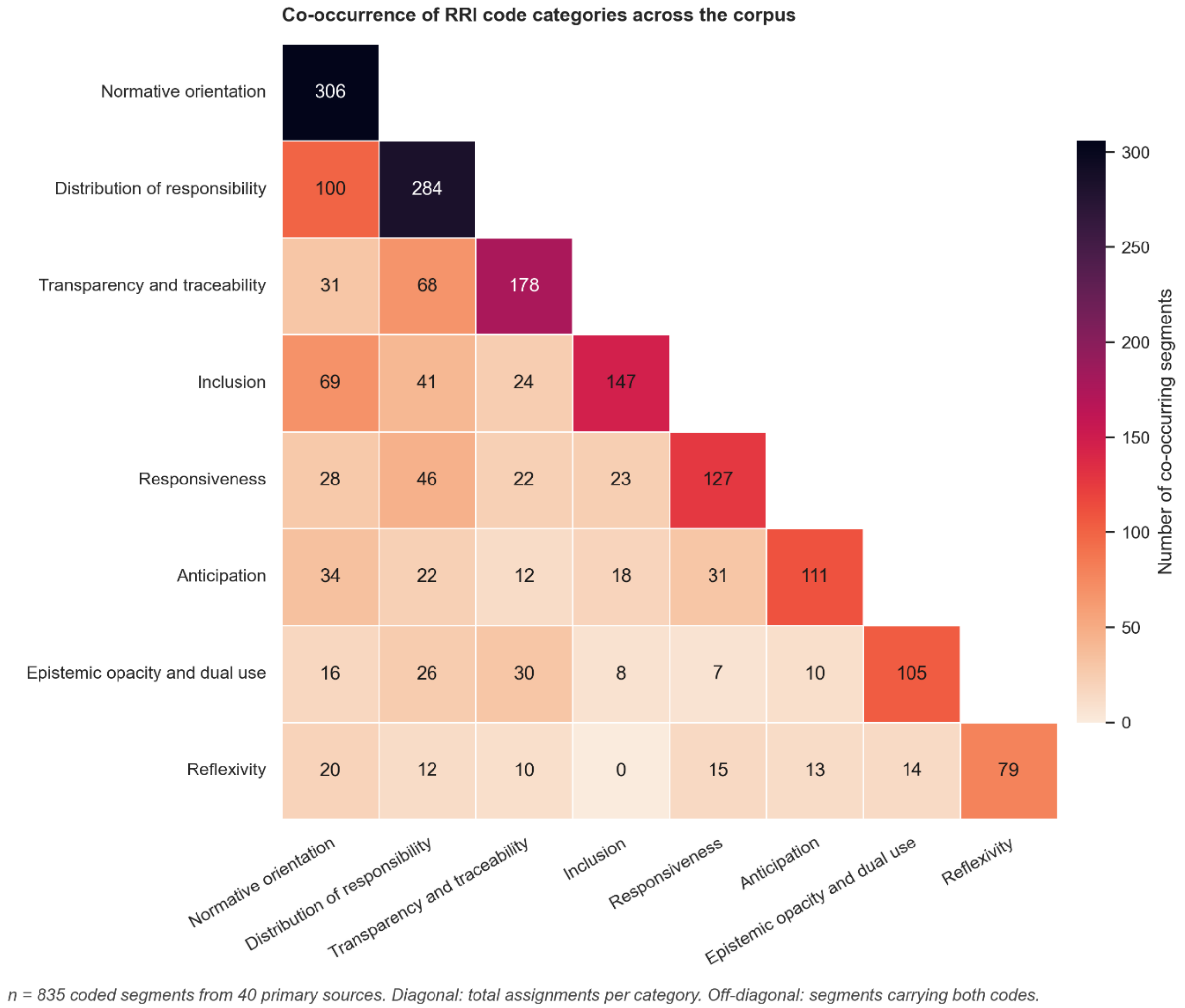

**Fig. 3.** Co-occurrence of RRI code categories.

### *4.2 Existing frameworks and author protection*

#### *The normative architecture of copyright*

Copyright governance documents in the generative AI period inherit a normative orientation that is long-established but newly contested. The 2026 UK Report on Copyright and Artificial Intelligence states that copyright law has long sought to balance the rights given to creators with the wider public benefit of information exchange, and the 2026 European Parliament Resolution insists that technological development and established rights "should not be mutually exclusive." The language of balance can obscure an asymmetry in bargaining power and access to evidence between individual authors and large model developers.

The DSM Directive (EU 2019/790) makes this asymmetry explicit: "Authors and performers tend to be in the weaker contractual position" when granting licences or transferring rights, and the individual author is identified as the subject most in need of protection. Lucchi (2025) argues that the opt-out architecture creates "structural risks" to the European creative economy when applied to large-scale commercial training.

The US Copyright Office, in its Part 3: Generative AI Training report, frames the same tension as one of "striking an effective balance" between technological innovation and a thriving creative community; in the US context the balance resolves more readily toward innovation, with authors' rights as one interest among several.

*Anticipation*

Existing frameworks show a consistent pattern of anticipation: an awareness that current law may be inadequate for AI development as it actually proceeds, combined with institutional reluctance to legislate prematurely. The UK Report on Copyright and Artificial Intelligence acknowledges uncertainty whether "reforms to copyright would secure significant investment" in AI development. Deferral preserves optionality while leaving authors unprotected in the interim.

At the intersection of deployment, inference, and autonomous agents, the UK Report concedes that "it is also unclear how copyright impacts the deployment of AI, particularly as the use of autonomous agents becomes more widespread"; such deployment-side phenomena are treated here as boundary evidence of the limits of input-side governance. *Thomson Reuters v. Ross* illustrates the doctrinal difficulty: since "there is no computer code whose underlying ideas can be reached only by copying their expression," prior reasoning from software cases does not transfer to AI training.

The UK Copyright and AI Impact Assessment extends this temporal problem, flagging that synthetic training data "could complicate the role of copyright works" in training the next generation of AI models, an uncertainty that receives no governance response; the GPAI Code of Practice commits its signatories to interim machine-readable reservation protocols "while anticipating the development of standards," codifying action under uncertainty as a feature of compliance.

*Reflexivity*

Across the corpus, instruments periodically acknowledge the inadequacy or unstated presuppositions of their own frameworks; this reflexivity is most visible where courts and legislators revisit prior positions.

*Thomson Reuters v. Ross* also illustrates reflexivity at the procedural level: the Delaware court revised its own 2023 summary judgment opinion, producing doctrinal instability within a single proceeding. The European Parliament Resolution is equally candid, its rapporteur's explanatory statement acknowledging that the DSM Article 4 exemption "was not drafted with the intention of enabling the use en masse of copyright-protected material" by generative AI; the instrument identifies its own drafting gap.

The UK Government demonstrates the same pattern institutionally. Having proposed a broad TDM exception with opt-out in the 2024 consultation, it withdrew the position: the exception "is no longer the government's preferred way forward." The Impact Assessment confirms that the reversal rests on an evidentiary gap, conceding that the evidence base is evolving rapidly and that it "does not establish a preferred option"; the result is institutional non-decision under genuine uncertainty rather than a failure of process.

A doctrinal contradiction within the Copyright, Designs and Patents Act 1988 (CDPA) reinforces the point: the consultation highlighted an "apparent contradiction between section 9(3) CDPA, which applies to original works without a human author, and the originality standard," associated with human qualities such as personality. Existing frameworks have not resolved this tension.

*Epistemic opacity and dual use*

A recurring structural condition in the corpus is that the actors whose cooperation is required to establish copyright compliance are also in exclusive possession of the information needed to assess it: only 22% of artists surveyed for the UK Impact Assessment reported that their work had been used for AI training, while 63% were unsure.

The UK Report documents how this opacity is maintained by law as well as practice: there is "no requirement in UK law for an AI developer or provider to publicly make available information about the copyright works used to develop an AI model," and providers are not required to disclose what "would compromise their trade secrets." Authors may therefore be unable to assess or substantiate a potential infringement outside litigation or regulatory processes.

In *GEMA v. OpenAI*, the court held that song lyrics memorised in model parameters were reproduced and that the TDM exception, which covers only preparatory copies for analysis, did not extend to them; memorisation was established by prompting the deployed system, not through any disclosure the developer was required to make.

The dual-use dimension is most acute at the training data level. Commenters to the US Copyright Office reported that "it is common for training datasets to include pirated works or works accessed by circumventing paywalls," and the Office notes that copyright-related information is sometimes deliberately removed from datasets. The complaint in *NYT v. Perplexity* alleges the same strategy at the inference stage: the Perplexity-User crawler "generally ignores robots.txt rules", and undeclared user agents allegedly impersonated Google Chrome when the declared crawler was blocked. The GPAI Code of Practice commits signatories not to circumvent access controls, but provides no independent means of verifying that commitment.

### *Inclusion*

The inclusion dimension of copyright governance concerns whose interests frameworks actually protect and who their design excludes; despite broad rhetorical consensus favouring individual authors, small publishers, and SME developers, the structural design of most proposed mechanisms disadvantages them.

The DSM Directive (recital 30 to Art. 8) acknowledges that "obtaining the prior authorisation of the individual rightholders can be very difficult" for out-of-commerce works, and (Art. 19) mandates transparency on "all revenues generated and remuneration due"; these provisions were designed for established licensing relationships, not mass, undifferentiated ingestion by training pipelines. The US Copyright Office predicts that smaller copyright owners "would either be overlooked in licensing deals or would receive substandard terms," a prediction confirmed by UK consultation findings that agreed remuneration "will be very small."

Opt-out mechanisms reproduce this exclusion at the level of implementation costs. The UK Impact Assessment warns of "a disproportionate effect on SMEs", since a reservation mechanism takes a greater share of an SME's total costs, and the UK Report is more direct: a TDM exception with opt-out "would disproportionately harm small creators, who are the least likely to reserve their rights". The European Parliament Resolution proposes the EUIPO as a trusted intermediary to manage exclusions in a way that is "simple and cost-efficient", identifying the inclusion deficit without supplying a structural remedy. The French Autorité de la concurrence's Decision 24-D-03 shows that these limits persist even for organised rightsholders: publishers wanting to opt out of AI training had to block Google's crawling of their content altogether, and Bard continued to draw on sites that had set the Google-Extended tag.

*Bartz v. Anthropic* approaches the same problem from a different direction: the court held that "a market for that use is not one the Copyright Act entitles Authors to exploit," foreclosing a remuneration claim not because authors are unharmed but because the statutory framework does not contemplate that market. The pirated-copies class subsequently settled for $1.5 billion (preliminary approval, September 2025), but the doctrinal point remains stable: in the US fair use context, inclusion is not an interpretive value.

*Constitutive silences in existing frameworks*

Five patterns above describe what the corpus addresses; the constitutive silences in the governance corpus (legislation, court decisions, soft law) are of equal analytical significance.

The UK Report notes that "authors, film directors and performers are granted moral rights under the CDPA," including the attribution and integrity rights, yet no instrument examined applies them to training-data practices. The corpus frames the copyright–AI problem almost exclusively in economic terms, so whether training and producing competitive outputs violates the integrity right, or whether removing authorship information from datasets violates the attribution right, remains unaddressed.

The TRAIN Act and CLEAR Act address copyright owners, but many working authors do not own their copyright, having transferred it to publishers, labels, or studios. Where rights have been transferred, licensing is ordinarily conducted by the transferee, and the governing contractual or statutory framework may provide no additional remuneration to the author.

Many of the most commercially significant models in deployment were trained on datasets assembled before any transparency, licensing, or opt-out regime existed, and the corpus consistently defers the resulting challenge. The EU Parliament opposes a global training licence in exchange for a flat-rate payment, calls for "a possible solution for the immediate, fair and proportionate remuneration for past uses," and the rapporteur's explanatory statement urges "an immediate, simple, flat-rate copyright fee for this use of 5 to 7% of global turnover". The AI Act's GPAI obligations apply to models placed on the Union market from 2 August 2025 onward, with the AI Act's transitional rules deferring obligations for earlier models until 2 August 2027. Thus, the coded corpus identifies no generally applicable binding redress mechanism for training uses completed before the relevant transparency, licensing, or opt-out regimes took effect. The Autorité's Decision 24-D-03 confirms the boundary from the enforcement side: even a binding remedy left past training in place, as Google stated that the opt-out "is only valid for the future" and already-used sites "cannot be removed from the corresponding models."

The TRAIN Act's subpoena mechanism, the CLEAR Act's proposed per-work notice requirement, and the UK's proposed input transparency measures identify which works were used, not how the model processed or retains them; the WIPO SCCR draft's note of "opacity in algorithmic royalties' allocation" as a governance gap marks a still deeper opacity within trained models.

The UK Impact Assessment notes that concentrated AI developer markets "can weaken the negotiation power of right holders" and that "most reported licensing deals are between large right holders and large AI developers"; it recognises the link between concentration of the market and weaker bargaining power of the creators but supplies no corrective mechanism. The UNESCO Recommendation calls for measures "to prevent abuse of dominant market positions, including by monopolies." The US National Policy Framework explicitly recommends against a regulatory response: "Congress should not create any new federal rulemaking body to regulate AI." The silence is constitutive: no new sector-specific regulator is contemplated.

### *4.3 Governance instruments and responsibility*

#### *Responsiveness*

Governance proposals across the corpus fall into three categories: transparency and disclosure duties, licensing market facilitation, and technical standards for rights reservation. The translation from diagnosis to operational instruments is, in most jurisdictions, incomplete. The UK Government's response to its own consultation is representative: it commits to "work with a range of industry and other experts to develop best practice on input transparency to help right holders assert their rights" and to "explore options for supporting human creativity and artistry," while explicitly declining to select a preferred regulatory option. The Data Use and Access Act 2025 institutionalises this deferral, requiring the Secretary of State to publish a report on AI and copyright within nine months and to "consider, and make proposals in relation to" technical standards, licensing, and enforcement, without itself creating any of those mechanisms. Across the corpus, responsiveness is more often a commitment to act than an act.

The CLA/ALCS/PLS Collective Licensing Scheme for AI Training (UK, announced in 2025, under development as of 2026) represents one of the few responsiveness instruments to have moved beyond mere commitment: it would assign remuneration duties to AI developers through a collective management organisation, partly addressing the individual opt-out problem by negotiating sector-wide licence terms on behalf of authors and publishers (*PLS and ALCS Agree to Development of Pioneering CLA Generative AI Licence*, 2026). The scheme is notable precisely because it addresses the inclusion gap identified above. Its limitation, however, is that it is still under development and its announced scope is confined to the rights managed by the UK collecting societies; training on the same works conducted outside UK jurisdiction would fall outside its reach.

The EU GPAI Code of Practice is the most operationally specific instrument in the corpus. Signatories commit to a copyright compliance policy with assigned internal responsibility, and are encouraged to publish a summary; they commit not to circumvent paywalls or technological access controls; and they commit to honouring robots.txt reservations and any other machine-readable rights reservations that have been "widely adopted by rightsholders" and "generally agreed through an inclusive process." However, the Code prescribes compliance on infrastructure that does not yet exist, namely agreed technical standards and lists of persistently infringing sites. The European Parliament Resolution acknowledges this directly, calling for the Code's guidelines to "be revised and treated as living documents requiring regular updates to address emerging challenges."

In the United States, the regulatory burden is displaced onto the courts. The National Policy Framework states that "the Administration believes that training of AI models on copyrighted material does not violate copyright laws" but "supports allowing the Courts to resolve this issue," recommending that Congress "should not take any actions that would impact the judiciary's resolution." The TRAIN Act and CLEAR Act propose transparency requirements and per-work notice obligations but leave liability to litigation. The US Copyright Office's Part 3 report reaches a consistent position: "licensing markets should continue to develop... In those areas where remaining gaps are unlikely to be filled, alternative approaches such as extended collective licensing should be considered to address any market failure." Responsiveness in the US instruments is contingent on market development and judicial resolution rather than regulatory design.

*Distribution of responsibility*

The corpus assigns copyright-governance duties across a chain of actors without consistently specifying what happens when those duties conflict or go unperformed. The GPAI Code of Practice extends its signatories' compliance obligations downstream: their duties apply "irrespective of whether a Signatory vertically integrates the model into its own AI system(s) or whether the model is provided to another entity based on contractual relations." The European Parliament Resolution reinforces this, insisting that transparency obligations cover not only developers of AI models, "as currently stipulated by Article 53 of the AI Act, but also by providers or deployers of AI systems."

Courts have begun to develop their own allocation logic. *Bartz v. Anthropic* holds that "not every person who merely intends to make a fair use of a work is thereby entitled to a full copy in the meantime, nor even to steal a copy so that achieving this fair use is especially simple or cost-effective," placing the burden of lawful acquisition on the developer rather than on dataset intermediaries. The same court found that "Anthropic later bought a copy of a book it earlier stole off the internet" but that subsequent purchase "will not absolve it of liability for the theft," rejecting post-hoc remediation as a responsibility-shifting mechanism. *In Re OpenAI* distributes responsibility differently: the court required OpenAI to produce a sample of 20 million de-identified chat logs while accepting OpenAI's own de-identification tool as an adequate privacy safeguard, effectively designating the developer as both the party possessing the evidence and the party authorised to sanitise it.

The UK Impact Assessment identifies the structural tension most directly: "the onus will be on AI service providers to ensure that they comply with the jurisdiction they are operating in," yet "under all options, SMEs will have less resources to determine whether the models they use are compliant with changes to copyright law and/or transparency measures." The corpus places primary responsibility on developers while acknowledging residual risk for downstream users, but no enforcement mechanism in the corpus reaches deployers.

*Transparency and traceability*

Transparency provisions are the most common governance instrument in the corpus, but they share a structural ceiling: what must be disclosed falls short of what enforcement would require. The EU Parliament Resolution calls for "an itemised list identifying each item of copyright-protected content used for training," applicable not only at training but "mutatis mutandis to any subsequent use of this content for other purposes, including inference, retrieval-augmented generation or fine-tuning." The TRUMP AMERICA AI Act proposes a requirement for AI developers to publish a Training Data Use Record and an Inference Data Use Record, with monthly updates for inference models, enforceable by the Federal Trade Commission (FTC). The GPAI Code of Practice commits its signatories to disclosure of web crawler identities and robots.txt configurations, and encourages publication of a copyright policy summary. The UK Competition and Markets Authority's Publisher Conduct Requirement extends transparency to the output side: Google must take reasonable steps to ensure that Search Content in search generative AI features is attributed "clearly, accurately," and must publish "Explanatory Information" explaining its approach to attribution. These extensions to inference and output are boundary cases for an input-side analysis: they mark the point at which transparency obligations begin to reach beyond the training corpus.

None of these instruments requires disclosure of model weights, training loss attributable to specific works, or any measure of a work's influence on model outputs. The WIPO SCCR draft identifies "opacity in algorithmic royalties' allocation" as a governance gap; the formulation accurately describes the symptom but not the underlying problem. *In Re OpenAI* illustrates the practical ceiling: the court ordered disclosure of output logs, not training data. The UK Report acknowledges that "metadata may be stripped from a digital file, which will remove creator and right holder information and preferences" and that "adherence to metadata, and its onward provision, relies on implementation by platforms and intermediaries." Transparency obligations, in other words, depend on a metadata infrastructure that is already broken at multiple points in the chain before any governance instrument can operate on it.

*Constitutive silences*

Three gaps define the outer boundary of what the governance instruments examined here can achieve.

The first is the *enforcement gap*. Proposed disclosure obligations across the corpus lack independent verification. The GPAI Code of Practice sets out a complaint mechanism through which rightsholders may submit "sufficiently precise and adequately substantiated complaints" to a designated contact point but requires no third-party audit and provides no sanction for non-compliance short of AI Act enforcement proceedings. The same pattern recurs across instruments: no independent audit or third-party verification (GPAI Code of Practice), no dedicated enforcement mechanism in the Act itself (California Assembly Bill 2013), enforcement limited to FTC fines (TRUMP AMERICA AI Act). The UK Competition and Markets Authority's Publisher Conduct Requirement, though legally binding, relies on confidential compliance reports and self-reported metrics for verification: the CMA will "require Google to submit a confidential version of the compliance report to the CMA with detailed metrics." The Autorité de la concurrence's Decision 24-D-03 shows the same problem inside a fully adjudicated monitoring regime: the Monitoring Trustee reported that "the remuneration offers transmitted are never accompanied by explanations enabling the Negotiating Parties to understand the calculation formula," and the Autorité held that where "the Monitoring Trustee is unable to verify a Commitment, this must be assessed as a noncompliance."

The second silence concerns *open-access and publicly funded research*. The DSM Directive's TDM exception (Art. 3) is designed for research institutions, not commercial AI developers; the Art. 4 commercial exception does not distinguish between a publisher's proprietary database and a publicly funded open-access repository. The TRAIN Act, CLEAR Act, and GPAI Code treat copyright-protected material as a uniform market commodity, without addressing works that are publicly funded, openly licensed, or published under Creative Commons terms. The coded dataset records the absence of any provision for open-access or CC-licensed works in the TDM context, publicly funded research, and scientific data as a distinct category.

Spichtinger (2026) demonstrates why this silence is particularly consequential for research data: Creative Commons licences predate large-scale commercial AI systems and contain no clause addressing computational reuse for AI training, and researchers who applied them "may not have anticipated that their work could be ingested by a machine learning system and repurposed to generate outputs." Where the training process qualifies as fair use in the United States, "CC-licensed work could be used for AI training without a legal requirement to follow the CC conditions," so CC BY's attribution requirement and CC BY-NC's non-commercial restriction alike can be lawfully disregarded (Spichtinger, 2026). This silence is directly relevant to this study's scope: scholarly publishing is disproportionately composed of such works, yet none of the instruments examined develops a distinct legal regime for them. Lucchi (2025) confirms this gap from the policy side, noting that the European Parliament study explicitly flagged the need for "harmonised opt-out mechanisms" and “equitable licensing models” but did not develop a distinct legal regime for openly licensed or publicly funded scholarly content.

The third silence is the absence of *binding international coordination*. The Indonesian SCCR/47/6 proposal calls for a legally binding international instrument on royalty governance, identifying challenges that “cannot be effectively addressed through isolated national measures,” and the WIPO SCCR/48/4 Draft Elements, framed as elements of a possible instrument, likewise identify “cross-border royalty flows” and “metadata fragmentation” as cross-border problems. The UK Impact Assessment acknowledges that any domestic transparency regime’s effectiveness depends on whether it applies to models trained abroad: “we assume for the purpose of this assessment that transparency measures may apply to models developed abroad, as well as those developed in the UK.” No binding instrument in the corpus resolves the question of which jurisdiction’s law governs training conducted outside any single territory. *Getty Images v. Stability AI*, the most densely coded document in the corpus, shows the same limit in litigation: the training claim was abandoned because there was no evidence that training occurred in the United Kingdom, and the court held only that the model weights were not an infringing copy, leaving the lawfulness of the training itself undecided.

The European Parliament Resolution is the only instrument in the corpus that addresses the jurisdictional gap directly. Recital AH asserts that “the principle of the territoriality of copyright protection must be adapted to the training of generative AI systems so as to ensure that the use of European content is subject to EU law even when such training takes place outside the EU,” and operative paragraph 16 calls for models that do not meet this requirement to be “barred from being placed or made available on the EU market.” The Resolution thus proposes a market-access condition as a functional substitute for extraterritorial jurisdiction: EU copyright rules would apply at the point of market entry regardless of where training occurred. The proposal carries no legal force in its current form; a parliamentary resolution addressed to the Commission does not itself create binding obligations, and the Commission has not yet initiated a corresponding legislative procedure. Whether a market-access mechanism of this design is compatible with WTO obligations under the TRIPS Agreement and with the territorial scope of the AI Act remains an open question. The CMA’s Publisher Conduct Requirement (2026) illustrates the same constraint in binding form: its attribution and grounding obligations are confined to generative AI services “shown to UK end users,” so the remedy’s territorial scope matches the UK market rather than the borderless scale of training.

## 5. Discussion

The short answer to RQ1 is that the EU and the US do not simply regulate at different speeds but rather produce diverse governance profiles. Table 2 summarises the corresponding legal architectures, with the UK, China, and the international soft-law instruments outlined as comparators.

Table 2. Qualitative legal-architecture synthesis by jurisdiction

| Dimension | European Union | United States | United Kingdom | China | International instruments |
|---|---|---|---|---|---|
| Governance mode | Statutory ex ante specification. | Ex post adjudication with declaratory policy. | Deliberative deferral pending evidence. | Administrative supervision within a security and development frame. | Non-binding coordination; a binding instrument only proposed. |
| Principal instruments | DSM Directive Arts. 3–4; AI Act Art. 53; GPAI Code of Practice. | Enacted state law (California AB 2013); fair use litigation; pending bills | Data Use and Access Act 2025 and related reports; CMA Publisher | Interim Measures for Generative AI Services; Global AI Governance Action Plan. | UNESCO Recommendation; OECD AI Principles; WIPO SCCR proposals. |
| | | (TRAIN, CLEAR); a discussion draft not introduced (TRUMP AMERICA AI Act). | Conduct Requirement. | | |
| Legal status | Binding statutes and regulation; the Code of Practice is soft law. | Judgments binding inter partes; bills pending. | Act in force; Report and Impact Assessment advisory; CMA requirement binding. | Interim Measures in force as a regulatory instrument; the Global AI Governance Action Plan is a policy commitment without legal force. | Soft law; proposals under discussion. |
| Enforcement locus | AI Act public supervisory regime. | Federal courts; FTC (proposed); no sectoral regulator. | CMA for the conduct requirement; Secretary of State reporting duty. | State supervision of providers. | None; duties addressed to states. |
| Verification and enforcement adequacy | Public supervision under the AI Act; Code of Practice complaints lack independent audit. | Litigation-dependent; discovery is slow and costly (In re OpenAI); no sectoral verifier. | Binding CMA requirement verified through confidential, self-reported metrics. | State supervision; no rightsholder-facing verification. | None; implementation is reviewed only among states. |
| Attribution | Corpus-level disclosure; itemised list proposed, not binding. | Public dataset-level summaries (AB 2013); work-level records to owners (TRAIN) or the Register (CLEAR), both pending; no author-facing duty. | No training-related duty; binding output-side attribution to publishers under the CMA requirement; CLA/ALCS/PLS collective licensing under development. | Labelling of generated content; no attribution of sources. | Metadata integrity as a precondition; no duty created. |
| Remuneration route | Collective licensing and fair compensation exist in the acquis but create no AI-specific duty. | No statutory route; court-supervised settlements only (Bartz v. Anthropic). | CLA/ALCS/PLS scheme under development; no statutory route. | None. | Royalty governance proposed only (SCCR/47/6; SCCR/48/4). |
| Major silence | No distinct regime for openly licensed or publicly funded scholarly works under Art. 4. | No statutory baseline for training; liability left to litigation. | No preferred regulatory option. | No mining exception, reservation mechanism, or remuneration route. | No binding cross-border coordination. |

Note. Qualitative synthesis by the author based on the instruments analysed in Sections 4.2 and 4.3; entries are not coded frequencies (see Table 1). Legal status distinguishes binding law, judgments, pending legislative proposals, and soft law.

No system in the corpus constructs attribution as an entitlement of the author for the use of works in training. In the EU, attribution is absorbed into corpus-level disclosure under Article 53(1)(d), and the European Parliament Resolution's demand for an itemised list extending "mutatis mutandis" to inference and retrieval-augmented generation carries no binding force. In the US, California AB 2013 requires public dataset-level summaries. The pending TRAIN Act would let legal or beneficial copyright owners subpoena records identifying their works, and the pending CLEAR Act would require per-work notices published in a public Copyright Office database. Neither proposal creates a general attribution entitlement based solely on authorship. In China, labelling duties attach to generated content rather than to its sources. In the international instruments, attribution appears only as a metadata-integrity precondition for royalty allocation.

These profiles are products of legal architecture rather than of differing levels of concern. Four architectures are represented: statutory ex ante specification (EU), ex post adjudication supplemented by declaratory policy (US), deliberative deferral pending an evidence base (UK), and administrative supervision within a security-and-development frame (China). The international layer adds a fifth mode, non-binding coordination, which generates declarations rather than compliance records.

RQ1 asked how US and EU instruments frame and operationalise the protection of authors whose works are used for LLM training. The coded corpus supports a four-part answer. First, both systems frame protection procedurally rather than as an entitlement. Second, the two systems operationalise that protection at different points in the chain. The EU acts ex ante through codified duties on providers, the US ex post through discovery and adjudication. Third, neither system operationalises attribution in the sense the RRI framework requires. Fourth, the protection offered by both systems depends on evidence that the instruments in the coded corpus do not generally require developers to provide at the work level. The UK and China vary the first three findings without contradicting them, and the international layer names the coordination problem that all four leave open.

Sag & Yu (2024) identify an emerging international equilibrium around text and data mining. The coded profiles suggest that convergence, where it exists, lies in shared silences rather than in shared instruments. The divergence reported by Chopra et al. (2025) and Carlini et al. (2026) is corroborated by the coding and can be specified: it lies in the timing and generality of obligations, not in the values invoked to justify them. Lucchi's (2025) finding of structural risk in opt-out architectures is supported by the inclusion coding, where opt-out costs fall disproportionately on small creators and SMEs.

Owen et al. (2021) concluded that RRI left industrial and commercial innovation largely unaddressed. The coding supports this: reflexivity is the least represented category corpus-wide (5.9%) and reaches its minimum in the US sub-corpus (2.1%). Woolley et al. (2024) argue that RRI travels badly beyond north-western Europe, and the comparator data refine rather than confirm that claim. The Chinese instruments record 38.0% of assignments to the four AIRR dimensions, close to the EU's 33.6% and above the US's 24.3%, through appeals such as "building an inclusive multi-stakeholder governance model." What does not travel is reflexivity, at 4.2% in the Chinese material and 2.1% in the US material. Trade-secret protection illustrates a barrier to verification; the crawler concealment alleged in *NYT v. Perplexity* would, if substantiated, illustrate strategic opacity. The temporal pattern is consistent with the Collingridge dilemma: operational governance instruments emerged mainly after generative models had already been deployed at scale.

Two constraints on the answers to RQ2a and RQ2b follow from these findings. An instrument capable of closing the gap must generate the evidence on which it depends rather than presuppose it, which points to measurement and disclosure duties defined at the level of the individual work. An allocation of responsibility must reach the actors who hold that evidence, which the corpus shows to be developers and intermediaries rather than the authors on whom the burden of proof currently falls.

## 6. Toward responsible copyright governance

RQ2a and RQ2b asked what governance instruments could address the gaps identified in Section 4 and what responsibilities the actors involved should assume. The design is normative rather than a direct output of the coding. Each proposed instrument responds to a documented governance gap and extends a mechanism observed in the corpus in incomplete, non-binding, or non-auditable form. The eight coded categories accordingly change their function: in Section 4 they served as evaluation criteria applied to existing law; here they serve as design requirements (Von Schomberg, 2011).

Two features carry the design's originality. The first is auditable work-level provenance (instruments 1 and 2): recording each ingested work, its rights basis, and the date of ingestion closes the evidentiary problem that makes litigation the only current verification mechanism: whether a work was used no longer depends on chat-log discovery, as in *In re OpenAI*. Because public work-level disclosure may expose confidential information, the record is held for audit by a qualified third party reporting to the supervisory authority, on the model of the monitoring trustee in the French competition proceedings; disclosure to authors and researchers remains aggregate, as under the GPAI Code of Practice. The second is the default author-facing remuneration share (instrument 3): extended collective licensing shifts the default from individual opt-out to collective licensing, and a defined share of what is collected is guaranteed to authors rather than left to intermediary discretion. The remaining instruments are specified in Table 3.

Table 3. Governance instruments mapped to the coded categories

| Instrument | Categories addressed | Mechanism | Closest corpus analogue |
|---|---|---|---|
| Work-level provenance and disclosure at ingestion | Transparency and traceability; normative orientation | Developers record each ingested work, its rights basis, and the date of ingestion; public disclosure remains aggregate | EP Resolution itemised-list demand; California AB 2013 training-data records; CLEAR Act per-work notice (pending) |
| Independently auditable training-data records | Transparency and traceability; responsiveness; reflexivity; epistemic opacity and dual use | The ingestion record is held so that a qualified third party can audit it without exposing trade secrets; findings feed supervision | GPAI Code of Practice complaint mechanism; CMA compliance reports (both unaudited) |
| Extended collective licensing with a default author-facing remuneration share | Distribution of responsibility; responsiveness; inclusion | Unreserved works are licensed through collecting societies; a defined share of collections is guaranteed to authors | DSM Directive Art. 12; CLA/ALCS/PLS scheme (under development) |
| Machine-readable rights-reservation standard | Normative orientation; inclusion; anticipation | A single machine-readable expression of reservation, honoured by crawlers at ingestion, drafted in a multi-jurisdiction process seating authors and developers | DSM Directive Art. 4 opt-out; robots.txt disclosure under the GPAI Code |
| Retroactive settlement for pre-regime training | Distribution of responsibility; normative orientation | Compensation for works used before any regime took effect, administered through collecting infrastructure | Bartz v. Anthropic class settlement |
| Market-access condition | Distribution of responsibility; transparency and traceability; epistemic opacity and dual use | Market entry conditional on evidenced compliance with instruments 1-4 regardless of where training occurred; rebuttable presumption where transparency is unmet | EP Resolution, operative para. 16 |

Table 4 answers RQ2b in the terms of the distribution-of-responsibility category: the allocation follows capacity rather than formal title. Developers hold the ingestion record because only they can create it; regulators define the audit standard and operate the market-access condition because only they can compel compliance; collecting societies distribute because they hold the infrastructure; publishers expose rights metadata because they control the publication record. Individual authors are assigned the duties they can discharge, setting reservations and participating in consultations, and are relieved of those they cannot, such as monitoring training corpora.

Table 4. Stakeholder responsibilities in the governance architecture

| Stakeholder | Responsibility in the governance architecture | Categories carried |
|---|---|---|
| Developers | Record ingested works at ingestion; honour reservations; submit records to audit | Transparency and traceability; normative orientation |
| Publishers | Expose machine-readable rights metadata; avoid licensing terms that pre-empt author reservations | Transparency and traceability; distribution of responsibility |
| Collecting societies | Administer extended licensing; guarantee the defined author share; publish distribution accounts | Responsiveness; distribution of responsibility |
| Funders | Condition grants on interoperable rights metadata and reservation capability | Inclusion; anticipation |
| Institutions | Maintain rights metadata for institutional outputs; negotiate collectively rather than institution by institution | Inclusion; distribution of responsibility |
| Regulators | Define the audit standard; operate the market-access condition; review instruments periodically with affected stakeholders, reassessing assumptions and distributional effects | Reflexivity; responsiveness |
| Individual authors | Set reservations; take part in consultations and standard-setting; report distribution irregularities | Inclusion; anticipation |

The design narrows the Collingridge dilemma without resolving it: instruments 1 and 2 make ingestion records inspectable while models can still be modified or retrained; instrument 4 enables prospective rights signalling before ingestion; instruments 3 and 5 create routes to remuneration or redress without requiring individual litigation; and instrument 6 aligns compliance incentives at market entry. Its limits are equally clear from the coding. Bargaining asymmetries between individual authors and model developers exceed what the copyright instruments examined can correct on their own, indicating a complementary role for competition, contract, and sector-specific regulation. The Autorité's experience with monitoring suggests that competition law carries part of the load; a market-access condition reaches models placed on one market, not training conducted everywhere. Responsible copyright governance, on this reading, is not a new copyright regime: it is the existing regime made auditable, remunerated, and inclusive, the three properties the coding showed it lacks.

## 7. Limitations

Several delimitations qualify the findings.

First, the corpus is restricted to copyright instruments; the GDPR, the Data Governance Act, the Data Act, and the Digital Services Act fall outside its scope, so the interaction between copyright and data-protection regimes is not captured.

Second, the study examines AI use from the input side only, and the findings do not extend to output-side issues such as the copyright status of AI-generated works or liability for infringing outputs.

Third, the corpus was coded by a single researcher. Illustrative coded passages are quoted in the Results and the full coded dataset is openly available, but no inter-coder reliability check or delayed recoding was performed; the category boundaries therefore reflect one informed reading of the material.

Fourth, the corpus is purposive rather than representative. Jurisdictional percentages describe coded assignments in this corpus, not the prevalence or quality of governance in the underlying legal systems. The Chinese sub-corpus (two documents) and the international group (six heterogeneous instruments) permit only indicative comparisons, and sub-corpus composition shapes the profiles: the UK shares for anticipation and reflexivity reflect the deliberative genre of the Report and Impact Assessment, while the US profile is dominated by litigation documents.

**Acknowledgements**

This study was initially presented at the EU-SPRI Annual Conference 2026 in Valencia, in the section "Artificial intelligence (AI) in science: critical perspectives from responsible innovation."

**Funding**

This publication has been supported by the RUDN University Development Program "Strategy-2030", project No. C05.1 "Implementation of international and domestic academic cooperation and enhancement of academic reputation."

**Disclosure statement**

No potential conflict of interest to be reported by the author.

**Data availability statement**

The codebook, coded dataset, and corpus documentation supporting this study are openly available on Zenodo at https://doi.org/10.5281/zenodo.20555375. Full references and source links for the documents analysed are provided in the accompanying supplement.

**AI use statement**

During manuscript preparation, the author used Perplexity AI in Computer mode to assist with proofreading and language correction. The author reviewed the suggested changes and takes full responsibility for the accuracy, integrity, and final content of the manuscript.